\documentclass[preprint,12pt]{elsarticle}

\usepackage{amssymb}
\usepackage{amsmath}
\usepackage{appendix}
\usepackage{graphicx}
\usepackage{siunitx}
\usepackage{tikz}
\usepackage{tikz-3dplot}
\usepackage{caption}
\usepackage{subcaption}
\usepackage{xcolor}

\newcommand{\bi}{\begin{itemize}}
\newcommand{\ei}{\end{itemize}}
\newcommand{\be}{\begin{equation}}
\newcommand{\ee}{\end{equation}}
\newcommand{\ba}{\begin{align}}
\newcommand{\ea}{\end{align}}

\newcommand\nc{\newcommand}
\def\p{\partial}
\nc\pad[2]{\frac{\p #1}{\p #2}} 
\nc\padd[2]{\frac{\p^2 #1}{\p
{#2}^2}} 
\nc\nd[2]{\frac{\text{d} #1}{\text{d} #2}} 
\nc\pat[2]{\frac{D #1}{D #2}} 
\nc\ov{\overline} 
\nc\degree{^{\circ}} 
\nc\ra{\rightarrow} 
\nc\Ra{\Rightarrow} 
\newcommand{\Da}{\mathrm{Da }}

\newcommand{\bea}{\begin{eqnarray}}
\newcommand{\eea}{\end{eqnarray}}
\newcommand{\beas}{\begin{eqnarray*}}
\newcommand{\eeas}{\end{eqnarray*}}
\newcommand{\Tm}{\mathrm{T}_m}

\newcommand{\ord}[1]{\mathcal{O}(#1)}

\begin{document}

\begin{frontmatter}



\title{Approximate solutions to the shrinking core model and their interpretation}


\author[inst1]{Cristian Moreno-Pulido}
\ead{cristian.moreno@udg}
\author[inst2,inst3]{Rachael Olwande}
\ead{rachael.olwande@upc.edu}
\author[inst2]{Tim Myers}
\ead{tmyers@crm.cat}
\author[inst3]{Francesc Font}
\ead{francesc.font@upc.edu}

\affiliation[inst1]{organization={Dept. Comp. Science, Appl. Maths and Stats, Universitat de Girona},
city={Girona},
country={Spain}
}
\affiliation[inst2]{organization={Centre de Recerca Matemàtica},
city={Bellaterra, Barcelona},
country={Spain}}
\affiliation[inst3]{organization={Dept. Fluid Mechanics, Universitat Politècnica de Catalunya},
city={Barcelona},
country={Spain}}

\begin{abstract}
The shrinking core model (SCM) describes the reaction of a solid particle with a surrounding fluid. In this work, we revisit the SCM by deriving it from the underlying physical processes and performing a careful non-dimensionalisation, which highlights the limitations of the commonly used pseudo-steady-state approximation, particularly in liquid-solid systems where fluid and solid densities are comparable. To address these limitations, we derive approximate analytical solutions using a perturbation method that improves upon the pseudo-steady-state model. We also obtain a small-time solution capturing early transient behaviour. A semi-implicit finite difference scheme is implemented to solve the full model numerically and benchmark the analytical approximations. We demonstrate that the perturbation solution provides significantly improved accuracy over the pseudo-steady-state model, especially in diffusion-limited regimes. Finally, we propose a simple fitting procedure combining the perturbation with the early-time solutions to estimate physical parameters from experimental data at minimal computational cost.
\end{abstract}



\begin{keyword}
Mathematical modelling \sep shrinking core model \sep perturbation methods \sep moving boundary \sep Stefan problem \sep gas-solid reactions
%%\PACS 0000 \sep 1111
%%\MSC 0000 \sep 1111
\end{keyword}

\end{frontmatter}


\section{Introduction}
\label{sec:introduction}

The shrinking core model (SCM) is an idealised mathematical model developed to describe the reaction between a solid particle and a surrounding reactant fluid. The reaction occurs at a sharp interface that moves inward into the solid interior as the fluid diffuses through the growing, inert, porous, product layer. It is a classic model that has been applied to a multitude of physical problems, including: oxidation \cite{diez2018determination}, corrosion \cite{sato1988corrosion}, mineral leaching \cite{crundwell1997mathematical}, anaerobic digestion \cite{da2013shrinking}, combustion \cite{Yagi55}, dehydration \cite{lan2015shrinking}, hydrolysis \cite{Yoshi01}, catalyst regeneration \cite{scott2016elements}, smelting \cite{LIU20142206}, glass fiber diffusion \cite{SARKAR20092874}, in  the capture of environmental contaminants and calcination (for example through batch adsorption \cite{AHN201713}). The present study is motivated by analysis of hydrogen storage in metal hydrides \cite{smith2001thermodynamics, oliva2018hydrogen, song2018numerical,li2020kinetics, cai2021rate, wang2021kinetics, wang2022simulation}, however, in the following we leave the analysis general such that it may describe liquid or gas interactions with a solid core.

Mathematical modelling of the SCM is generally based on the pseudo-steady state (PSS) approximation. The physical basis for the PSS is that the conversion process is much slower than the diffusion of mass to the reaction front. Mathematically this permits the neglect of the time derivative in the concentration diffusion equation, hence the concentration satisfies a steady-state equation. However, since the boundary is actually slowly moving, it is not a true steady-state, as the time-dependent boundary position is present in the solution. The PSS is a well-known approximation applied in a variety of fields, such as phase change, enzyme kinetics, lubrication theory (where the time derivative is dropped from the Navier-Stokes equations) or the related thin film theory, see \cite{Bisc63, Fowler, Myers98} for example. The problem studied in the present paper is analogous to a spherically symmetric one-phase change (Stefan problem) with a variable phase change temperature. This has been studied extensively in the analysis of nanoscale melting and nanocrystal growth \cite{Font2013,Font15,Ribera16,Myers19,Myers20}.
The classical SCM is often attributed to Yagi and Kunii \cite{Yagi55} and is described in detail in \cite{Froment, Levenspiel,Szekely}.

The applicability of the PSS has been questioned by a number of authors. Levenspiel \cite{Levenspiel} states that his analysis is based on the assumption that the surrounding fluid is a gas, but goes on to assert that it is equally applicable to a liquid. Wen \cite{Wen68} clearly states that the PSS is valid for most solid-gas reaction systems but not for solid-liquid systems, unless the liquid reactant concentration is very low.
Lidell \cite{liddell2005shrinking} focusses on reminding readers of the approximations and limits of the original models, which appear to have been lost over time and, in particular, on the range of applicability of the PSS assumption. They summarise Wen's result \cite{Wen68} stating that the non-dimensional coefficient of the time derivative in the diffusion equation must be less than 0.1 (which leads to errors of the order 10\%) to avoid erroneous results (in the following we will derive a similar limitation). A key result is that they point out the different system behaviour at small times (something we will also discuss later in this paper). 

There are clearly situations where the non-dimensional  coefficient of the time derivative in the diffusion equation is not negligible, for example at early times (hence the different behaviour mentioned in \cite{liddell2005shrinking}) or with liquid-solid systems. In these cases some form of correction must be sought. One attempt was provided by Theofanous and Lim \cite{THEOFANOUS19711297}. After changing all the variables in the problem, they integrate the heat equation twice to determine an integral equation for the concentration. This permits a solution to be obtained by repeated integration (the idea being that with each further integration the solution becomes more accurate). However, the method requires an initial solution which their approach does not provide, hence they apply a PSS approximation. To make for a simpler solution form they have imposed constant concentration at either boundary which results in a linear concentration. When this is placed into the integral formulation a cubic equation is obtained (their 2nd order solution). Replacing this back into the integral the 3rd order, obviously, is a quintic. The process could be carried further following a standard iterative  procedure but at the quintic stage the algebra is already becoming very cumbersome.
However, there is no proof that this iterative method converges on the correct solution. Moreover, the constant concentration boundary condition corresponds to the diffusion dominated case  described in Levenspiel \cite{Levenspiel} but the diffusion-controlled PSS is a cubic equation: their starting choice corresponds to the kinetic-controlled limit (we will demonstrate this more clearly in subsequent sections). The solution in \cite{THEOFANOUS19711297} involving the sum of their first and second order approximations, which is a cubic, may then be viewed as a somewhat inconsistent approximation to the standard model involving both kinetic and diffusion mechanisms - their governing equations neglect the kinetic reaction but their starting solution is equivalent to the kinetic solution. Liddell  \cite{liddell2005shrinking} states that the full cubic form may be obtained through elementary calculus. 

A review of different solid-state kinetic models can be found in \cite{khawam2006solid}, where a plethora of different models are briefly mathematically motivated and classified into one of the following categories: nucleation, geometrical contraction, diffusion, and those models described by the order of the underlying chemical reaction. 
A critical evaluation of the SCM can be found in \cite{gbor2004critical}. In \cite{gbor2004critical}, the model is extended to multiple particles, and the authors investigate the role of the particle size distribution and the effects of neglecting it. The main conclusion is that if the distribution is wide, with a large variance, the kinetic equations can be affected in a non-negligible way, as supported by experimental data. The interplay between multiple evolving particles is well-documented in the case of crystal growth, through the process of Ostwald ripening. A mathematical model equivalent to the SCM, but involving multiple spherical crystals is described in \cite{Fane21} to show that smaller particles may grow and then decay as larger particles use up the material available in the surrounding solution.

The paper is organized as follows. In Section 2, we derive the shrinking core model, non-dimensionalize it, and discuss the distinct time scales involved in the reaction process. We also revisit the widely used PSS approximation, examining its derivation and highlighting the limitations of several common simplifications. In Section 3, we develop approximate analytical solutions using the perturbation method, thereby providing corrections to the PSS forms. Additionally, we analyze the small-time limit and derive explicit expressions describing the solution behaviour in this regime. In Section 4, we formulate a semi-implicit finite difference numerical scheme to solve the full model. In Section 5, we compare and discuss the various analytical and numerical solutions found. We also illustrate how the analytical solutions can be used to infer physical parameter values from experimental data. Finally, Section 6 presents our conclusions.

\section{The Shrinking Core Model}
\label{sec:MathModel}

The SCM is an idealised model to describe the reaction and evolution of a solid particle with a surrounding fluid.
In a typical scenario, the fluid reacts with the solid, leading to a shrinking solid core surrounded by a growing outer porous layer. The outer layer is referred to as the $\beta$-phase and the inner solid core as the $\alpha$-phase. Fluid diffusing through the  $\beta$-phase, allows the reaction at the outer surface of the $\alpha$-phase to continue until the solid core is fully consumed, as illustrated in Fig.~\ref{fig:core_shell_evolution}. The model is based on the assumption that the interface separating the $\alpha$-phase from the $\beta$-phase, where the reaction takes place, is narrow, such that it may be treated as a sharp interface, $r^*=s^*(t^*)$.

\begin{figure}[htbp]
    \centering
    \begin{minipage}[t]{0.48\textwidth}
        \centering
        \begin{tikzpicture}[scale=0.5]
        % Thin Red Shell
        \shade[ball color = gray!30, opacity = 0.9] (0,0) circle (2.1cm);

        % alpha phase
        \shade[ball color = gray!30, opacity = 0.5] (0,0) circle (2cm);
        \draw (0,0) circle (2cm);
        \draw (-2,0) arc (180:360:2 and 0.6);
        \draw[dashed] (2,0) arc (0:180:2 and 0.6);
        \fill[fill=black] (0,0) circle (1pt);
        \draw[solid] (0,0 ) --   (-2,0);
        \node at (0,0.9) {\scriptsize $\alpha$-phase};

        % beta phase
        \shade[ball color = green!20, opacity = 0.3] (0,0) circle (3cm);
        \draw (0,0) circle (3cm);
        \draw (-3,0) arc (180:360:3 and 0.6);
        \draw[dashed] (3,0) arc (0:180:3 and 0.6);
        \fill[fill=black] (0,0) circle (1pt);
        \draw[solid] (0,0 )  (3,0);
        \node at (0,2.3) {\scriptsize $\beta$-phase};

        % Inwards arrows
        \foreach \angle in {30, 60, 90, 120, 150, 210, 240, 270, 300, 330} {
          \draw[->, thick] ({2*cos(\angle)}, {2*sin(\angle)}) -- ({1.6*cos(\angle)}, {1.6*sin(\angle)});
        }

        % Partícles around the sphere
        \foreach \x/\y in {
          3.5/2-3.6/0.7, 3.2/-1.0, -3.3/-1.3,  -3.9/1.0,
          2.8/2.6, -2.9/2.4, 3.0/-2.7, -3.0/-2.8,   -4.1/-0.5,
          0.5/3.5, -0.7/3.6, 1.0/3.3, -1.2/3.4, 2.0/3.8, -2.3/3.9,
          0.5/-3.5, -0.8/-3.6, 1.3/-3.2, -1.1/-3.5, 2.4/-3.9, -2.5/-3.7, 3.0/1.5, -3.1/1.6, 2.5/2.5, -2.6/2.6, 3.2/0.0, -3.3/0.1,
          0.0/3.3, -0.5/3.1, 1.8/2.7, -2.0/2.6, 2.9/-1.2, -3.0/-1.3,-1.1/3.5, 2.4/3.9, -2.5/3.7, 3.0/-1.5, -3.1/-1.6, 2.5/-2.5, -2.6/-2.6, -3.2/0.0, -3.3/-0.1,
          0.0/-3.3, -0.5/-3.1, 1.8/-2.7, -2.0/-2.6, 2.9/1.2, -3.0/1.3
        } {
          \fill[blue!70] (\x,\y) circle (2pt);
        }

        % adsorbed particles exterior
        \foreach \angle in {47, 129, 283, 6, 198, 312, 74, 355, 225, 163} {
          \fill[blue!90!black] ({3*cos(\angle)}, {3*sin(\angle)}) circle (2pt);
        }

        % diffusion of particles
        \foreach \angle in {47, 129, 283, 6, 198, 312, 74, 355, 225, 163} {
          \draw[->, dashed, red!80!black] 
            ({3*cos(\angle)}, {3*sin(\angle)}) 
            -- ({2.2*cos(\angle)}, {2.2*sin(\angle)});
          \fill[red!80] ({2.6*cos(\angle)}, {2.6*sin(\angle)}) circle (1.5pt);
        }
      \end{tikzpicture}
        \caption*{(a)}
    \end{minipage}
    \hfill
    \begin{minipage}[t]{0.48\textwidth}
        \centering
        \begin{tikzpicture}[scale=0.5, every node/.style={font=\small}]
  % Define radii
  \def\R{1.2}  % Outer radius
  \def\gap{3.2} % Horizontal gap between spheres

  % Step data: {step, inner radius, outer color, inner color}
  \foreach \step/\rin/\outcol/\incol in {
    0/1.2/green!10/gray!30, 
    1/0.8/green!10/gray!30, 
    2/0.4/green!10/gray!30, 
    3/0.0/green!10/gray!30
  } {
    % X offset for this step
    \pgfmathsetmacro{\x}{\step*\gap}
    
    % Outer sphere (ash layer)
    \shade[ball color=\outcol] (\x,0) circle (\R);

    % Inner core (unreacted solid)
    \ifdim\rin pt>0pt
      \shade[ball color=\incol] (\x,0) circle (\rin);
    \fi

    % Labels: only for first and last step
    \ifnum\step=0
      \node at (\x,-1.9) {\( t = 0 \)};
    \else\ifnum\step=3
      \node at (\x,-1.9) {\( t = t_f \)};
    \fi\fi
  }

  % Time arrows between stages
  \foreach \i in {0,1,2} {
    \pgfmathsetmacro{\xstart}{\i*\gap+\R+0.1}
    \pgfmathsetmacro{\xend}{(\i+1)*\gap-\R-0.1}
    \draw[->, thick] (\xstart,0) -- (\xend,0);
  }
\end{tikzpicture}
        \caption*{(b)}
    \end{minipage}
    \caption{(a) Schematic of the shrinking particle showing external adsorption, internal diffusion, and reaction at the receding interface. (b) Temporal evolution of the shrinking core during the reaction process.}
    \label{fig:core_shell_evolution}
\end{figure}

Assuming spherical symmetry, the fluid concentration $c^*$ [mol/m$^3$] in the $\beta$-phase may be described by 
\begin{equation}
    \label{eq:SCM-DiffEq}
    \frac{\partial c^*}{\partial t^*} = \frac{D}{{r^*}^2} \frac{\partial }{\partial r^*}\left({r^*}^2 \frac{\partial c^*}{\partial r^*}\right)\,,\qquad \text{in}\qquad s^*(t^*) < r^* < R^*,
\end{equation}
where $D$ [m$^2$/s] is the diffusion coefficient and $s^*(t^*)$ is the position of the $\alpha$-phase/$\beta$-phase interface. The star superscripts indicate dimensional variables.  The outer surface is fixed at position $r^*=R^*$; in reality, the outer surface may grow due to the different densities of the phases, $\rho_{\alpha} \ne \rho_{\beta}$. Hence the constant $R^*$  must be considered as an approximation where  $\rho_{\alpha} \approx \rho_{\beta}$, such that the change in $R^*$ is negligible. 
An analogous problem \cite{Font15}, related to the spherically symmetric melting of gold nanoparticles, demonstrates that the change in radius is equal to $(\rho_{\beta}/\rho_{\alpha})^{1/3}$. For the case studied in \cite{Font15} this results in an approximately 5\% increase in outer radius so suggesting that, as a first approximation, we may neglect the density change. 

On the outer surface, the particle is in contact with the surrounding fluid. The ideal gas law gives
\begin{equation}
    \label{eq:SCM-BoundaryCond1}
    c^*(R^*,t^*) = \frac{P^*}{R_g T^*} \equiv c_R^*(t^*)\,,
\end{equation}
where $P^*$ and $T^*$ represent the pressure and temperature of the surrounding fluid, and $R_g$ is the ideal gas constant. In principle, $P^*$ and $T^*$ may vary with time. For now, we will assume they are constant, and hence $c_R^*$ is also constant. This is a usual setup in experiments on metal hydrides (see, e.g., \cite{wang2022simulation}).

The fluid is transported by diffusion from the outer boundary $r^* = R^*$ to the interface $r^* = s^{*+}(t^*)$, where the $+$ superscript indicates the position immediately adjacent to the boundary in the region occupied by $c^*$ (the superscript $-$ refers to the adjacent point in the $\alpha$-phase). Using Fick’s law, the mass rate reaching the interface may be written as
\begin{align}
    J &= 4 \pi (s^{*+})^2 D \left. \pad {c^*}{ r^*} \right|_{r^*=s^{*+}} \, .
\end{align}
This must match the flux caused by the conversion at the interface, which is assumed to be proportional to the difference in concentration across the interface:
\begin{align}
\label{Jk}
    J &= 4 \pi (s^{*+})^2 k (c^*(s^{*+},t^*) - c^*(s^{*-},t^*)) \, .
\end{align}
When dealing with the evolution of a solid interface, the fluid concentration at the boundary is unknown. The solid concentration may be expressed through the Ostwald–Freundlich relation:
\begin{align}
    c^*(s^{*-},t^*) = c^*_{\rm eq} \exp\left(\frac{\alpha}{s^*}\right) \, ,
\end{align}
where $\alpha$ is the capillary length. Typically, the capillary length is of the order of nanometres, and so we may assume $s^*(t^*) \gg \alpha$ and hence $c^*(s^{*-},t^*) \approx c^*_{\rm eq} = P_{\rm eq}/(R_g T)$ is the equilibrium concentration, and $P_{\rm eq}(T)$ is the threshold pressure for the reaction to take place.

Through equating the flux expressions we can now define the fluid concentration at the interface,
\begin{align}
    c^*(s^{*+},t^*) = c^*(s^{*-},t^*)+\frac{D}{k}\left. \frac{\partial c^*}{\partial r^*} \right|_{r^* = s^{*+}}  \approx c^*_{\rm eq}+\frac{D}{k}\left. \frac{\partial c^*}{\partial r^*} \right|_{r^* = s^{*+}} \, .
\end{align}
Since we solve the system only in the $\beta$-phase we now drop the $+$ notation such that $s^{*+} \ra s^*$ and define the concentration boundary conditions as 
\begin{align}
c^*(R^*,t^*)&=c^*_R\,, \qquad   
D \left. \frac{\partial c^*}{\partial r^*} \right|_{r^* = s^{*}} =  k(c^*(s^{*},t^*) - c^*_{\rm eq})  \,.
\end{align}

The diffusion problem is defined over an unknown domain, $r^* \in [s^*(t^*), R^*]$. The position $s^*$ may be  determined through a mass balance - the rate of mass gain for the $\beta$-phase must equal the mass rate of fluid crossing the moving boundary,
\begin{align}\label{mass_balance1}
\left.\frac{d m_{\beta}^*}{dt^*} = 4\pi s^{*2} M_{g} D \frac{\partial c^*}{\partial r^*}\right|_{r^* = s^*} \, 
\end{align}
where $m_{\beta}^*(t^*)$ is the mass of the $\beta$-phase and $M_g$ the molar mass of the fluid. Expressing the mass of the $\beta$-phase in terms of its density and volume  we obtain 
\begin{align}\label{mass_beta}
m_{\beta}^*(t^*) = \rho_{\beta} V_{\beta} = \rho_{\beta}\left(\frac{4}{3}\pi R^{*3}-\frac{4}{3}\pi s^{*3}\right). 
\end{align}
Combining Eqs. \eqref{mass_balance1} and \eqref{mass_beta} leads to 
\begin{align}
\frac{\rho_{\beta}}{M_g}\,\frac{ds^*}{dt^*} &= -\left. D\frac{\partial c^*}{\partial r^*}\right|_{r^*=s^*}\,. 
\end{align}
This defines the evolution of the moving boundary, where at $t^*=0$ the interface is at the particle boundary $s^*(0)=R^*$.

In summary, the final set of equations of the SCM reads  
\begin{align}
\frac{\partial c^*}{\partial t^*} &= \frac{D}{r^{*2}}\frac{\partial }{\partial r^*}\left(r^{*2}\frac{\partial c^*}{\partial r^*}\right)\,, \label{dim_prob_1}\\
c^*(R^*,t^*) &= c_R^*\,,\\
\left. D\frac{\partial c^*}{\partial r^*}\right|_{r^*=s^*}&= k\,(c^*(s^*,t^*)-c_{\rm eq}^*)\,, \label{dim_prob_3}\\
\frac{\rho_{\beta}}{M_g}\,\frac{ds^*}{dt^*} &= -\left. D\frac{\partial c^*}{\partial r^*}\right|_{r^*=s^*}\,. \label{dim_prob_4}
\end{align}
The system \eqref{dim_prob_1}-\eqref{dim_prob_4} involves a diffusion equation for concentration  over an evolving domain, as such it is a form of one-phase Stefan problem. Note, no initial condition is specified for the concentration, since the domain occupied by the fluid does not exist at time $t^* = 0$.

This system holds under the following assumptions:
\begin{enumerate}
    \item The particles remain approximately spherical throughout the process;
    
    \item The concentration at the outer boundary is replenished much faster than the diffusion process, such that $c_R^*$ is constant for all time;
    
    \item The density of the $\alpha$ and $\beta$ phases is approximately the same $\rho_{\alpha}\approx \rho_{\beta}$, such that $R^*$ is constant;

    \item The reaction occurs over a narrow region, such that it may be treated as a sharp interface defined by $r^*=s^*(t^*)$.

    \item Particles are sufficiently large such that $s^* \gg \alpha$, where $\alpha$ is the capillary length.
    
\end{enumerate}

It is common practice to work in terms of the reacted fraction, this is defined through the ratio of  the  reacted volume to the total volume of the particle,
\begin{equation}
    \label{eq:SCM-ReactedFraction}
    X(t^*)=1-\frac{s^{*3}}{R^{*3}}\, ,
\end{equation}
such that $X(0)=0$.

\subsection{Non-dimensionalisation}
\label{subsec:nondim}

We now define the following dimensionless variables,
\begin{equation}
    \label{eq:nd-quantities}
    r\equiv\frac{r^*}{R^*}\,, \qquad s \equiv \frac{s^*}{R^*}\,, \qquad t \equiv \frac{t^*}{\tau}\,, \qquad c \equiv \frac{c^*-c_{\rm eq}^*}{c_{R}^*-c_{\rm eq}^*}\, ,
\end{equation}
such that the system may be written
\begin{align}
\frac{R^{*2}}{D \tau}\frac{\partial c}{\partial t} &= \frac{1}{r^2}\frac{\partial}{\partial r}\left(r^2\frac{\partial c}{\partial r}\right)\,,\qquad s < r < 1\\ 
c(1,t)&=1\,,\\  
\label{crND1}
\left. \frac{\partial c}{\partial r} \right|_{r = s} &= \frac{k R^*}{D} \, c(s,t)\,,\\ %\label{eq:nd-BoundaryCond2}
\frac{ R^{*2} \rho_{\beta}}{\tau  M_g  (c_{R,0}^*-c_{\rm eq}^*) D}\frac{d  s}{d  t}&=- \left. \frac{\partial c}{\partial r} \right|_{r = s}\,   , \qquad s(0)=1.
\end{align}
The coefficients of the time derivatives suggest two possible time-scales, the standard diffusion time-scale $\tau = \tau_D = R^{*2}/D$ and the growth time-scale $\tau = \tau_g = \rho_{\beta} R^{*2} / (M_g (c_{R,0}^*-c_{\rm eq}^*) D)$. A third time-scale, the kinetic time-scale may be identified from the boundary condition at $r=s$: if we write $D=R^{*2}/\tau_D$ then the coefficient in \eqref{crND1} becomes $k \tau_D/R^*$. The diffusion time-scale quantifies the time taken for mass to diffuse to the interface, while the kinetic time-scale $\tau_k = R^*/k$ quantifies the rate at which it reacts there.

The non-dimensional system may now be expressed as
\begin{align}
\label{ctND}\frac{\tau_D}{\tau}\frac{\partial c}{\partial t} &= \frac{1}{r^2}\frac{\partial}{\partial r}\left(r^2\frac{\partial c}{\partial r}\right)\,, \qquad s < r < 1\\ 
c(1,t)&=c_R (t)\,,\\  
\label{crND}
\left. \frac{\partial c}{\partial r} \right|_{r = s} &= \frac{\tau_D}{\tau_k} \, c(s,t)\,,\\ %\label{eq:nd-BoundaryCond2}
\label{stND} \frac{ \tau_g}{\tau}\frac{d  s}{d  t}&=- \left. \frac{\partial c}{\partial r} \right|_{r = s}\,  \qquad s(0)=1.
\end{align}

The choice of $\tau$ may be motivated by the domain of interest - do we wish to examine, growth, diffusion or kinetic behaviour? The classic pseudo-steady-state solution follows from setting $\tau= \tau_g$, that is, we work on the time-scale of the evolution and require the coefficient of the time derivative in the diffusion equation $\Da = \tau_D/\tau_g \ll 1$, where $\Da$ is a form of Damk\"{o}hler number representing the ratio of reaction to diffusion rates (or equivalently diffusion to reaction time-scales). This agrees with often quoted restrictions for the PSS approximation, see \cite{Wen68} for example. However, it requires a further restriction which is not discussed elsewhere - the time-scale $\tau_g$ is chosen to permit analysis with a moving boundary.
Noting that the gradient at the interface is proportional to the concentration we may write the condition \eqref{stND} as 
\begin{align}
 \frac{d  s}{d  t}&=- \frac{\tau_D}{\tau_k} \, c(s,t) \, .
\end{align}
Consequently, the PSS can only hold if $\tau_D/\tau_k \gg \Da$ or $\tau_g \gg \tau_k$, i.e. the growth time-scale is much larger than that of the reaction.

Alternatively, if we choose to focus on early time behaviour we could set $\tau = \tau_D$ or even $\tau \ll \tau_D$ such that the time derivative cannot be neglected. This demonstrates that for sufficiently  early times there is no PSS approximation and the system behaviour may be markedly different to that at large times - {the PSS cannot be applied  for times such that}  $t = \ord{\tau_D} = \ord{R^{*2}/D}$. 

To focus on the PSS approximation we now choose $\tau =   \tau_g$. This leads to
\begin{align}
 \Da \frac{\partial c}{\partial t} &= \frac{1}{r^2}\frac{\partial}{\partial r}\left(r^2\frac{\partial c}{\partial r}\right)\,, \qquad s < r < 1 \label{nondim1}\\ 
c(1,t)&=c_R (t)\,, \label{nondim2}\\  
\left. \frac{\partial c}{\partial r} \right|_{r = s} &= \Tm \, c(s,t)\,, \label{nondim3}\\
 \frac{d  s}{d  t}&=-  \Tm c(s,t) \,  \qquad s(0)=1  \label{nondim4}\, ,
\end{align}
where the dimensionless parameters are
\begin{equation}
\label{eq:nd-timescales}
  \Tm \equiv \frac{R^* k}{D} = \tau_D/\tau_k \,, \qquad \Da \equiv \frac{M_g  (c_{R,0}^*-c_{\rm eq}^*) }{\rho_{\beta}} = \tau_D/\tau_g
    \, ,
\end{equation}
and $\Tm $, the ratio of  diffusion to reaction time-scales, is related to the Thiele modulus $h_T$  (such that $\Tm = h_T^2$). The Damk\"{o}hler number  $\Da=\tau_D/\tau_g$ is the ratio of the diffusive to the growth time-scale. The denominator represents the density of the solid $\beta$-phase, while the numerator is the density of the fluid. Wen's \cite{Wen68} previously stated restrictions for the PSS to hold, based on the relative densities of the phases may be viewed as stating $\Da \ll 1$.
In gas-solid systems $\Da \ll 1$. However, in liquid-solid systems it is not always the case that $\Da$ will be small and so the size of dimensional numbers should  be verified when applying the models.

Finally, the non-dimensional expression for the reacted volume is
\begin{equation}\label{reac_frac}
X(t) = 1-s^3 \, , \qquad X(0)=0 \, .    
\end{equation}

%%%%%%%%%%%%%%%%%%%%%%%%%%%%%%%%%%%%%%%%%%%

\subsection{The standard Pseudo-Steady State  model}
\label{PSSmodel}

The simplest manner to deal with the SCM system is to apply the pseudo-steady-state approximation. This is based on setting $\Da \ll 1$, so that the time derivative may be neglected in the diffusion equation. In terms of chemical reactions it is explained in \cite{Myer24} that the PSS hypothesis typically holds when a process involves
two or more distinct time scales. In the present case diffusion is much more rapid than the shrinkage rate of the core. Henceforth we will take $c_R(t)=1$, assuming there is a large supply of reactant outside of the sphere or it is constantly introduced into the system.

The integration of the resultant ordinary differential equation is then simple and, after applying the boundary conditions, leads to
\begin{align}
\label{PSSC}
    c = 1+B - \frac{B}{r} \, , \quad \mbox{where} \quad B = \frac{\Tm s^2}{1+\Tm s- \Tm s^2} = \frac{s^2}{\Tm^{-1} + s- s^2} > 0 \, .
\end{align}
After substituting for $c(s,t)$ the mass balance may  be written
\begin{align}
\label{PSSst}
    \nd{s}{t} =   \frac{-1}{\Tm^{-1} + s- s^2}  \, < 0 \, .
\end{align}
Integrating and applying $s_0(0)=1$ leads to
\begin{align}
\label{PSSSol}
     \frac{s^3-1}{3} - \frac{s^2-1}{2} -  \Tm^{-1} (s -1) =    t \, .
\end{align}
A similar cubic equation is derived, with a different scaling, in \cite{melchiori2014improving}.

The time until all of the $\alpha$-phase is used up, $t_f$, is found by setting $s(t_f)=0$, and so
\begin{align}\label{final_t_PSS}
    t_f =  \frac{1}{6} + \Tm^{-1} \, .
\end{align}

Various versions of the PSS solution may be found throughout the literature. Based on physical arguments Levenspiel \cite{Levenspiel} defines three forms of model, controlled by: diffusion through the gas film; diffusion through the ash layer; chemical reaction. The first of these we neglect, assuming the fluid concentration surrounding the sphere remains constant throughout the process. The second refers to diffusion through the $\beta$-phase controlling the process. The third is controlled by the reaction at the interface. These may be thought of in terms of the parameter $\Tm= \tau_D/\tau_k$. If kinetics controls the process, i.e. the reaction is much slower than the diffusion rate, $\tau_k \gg \tau_D$,  then 
the \textit{Kinetic-Controlled} result of Levenspiel \cite[Ch. 25.2]{Levenspiel}  comes from setting $\Tm \ll 1$ in \eqref{PSSSol} so that
\begin{align}\label{KinCon}
s \approx 1- \Tm t \, .
\end{align}
The \textit{Diffusion-Controlled} result, $\tau_D \gg \tau_k$, follows from the large $\Tm$ limit
\begin{align}
\label{DiffCon}
    \frac{s^3-1}{3} - \frac{s^2-1}{2} =    t  \, .
\end{align}

These two limiting cases, widely accepted by the community, must be treated with caution. Firstly, the PSS approximation is based on the assumption $\Da = {M_g  (c_{R,0}^*-c_{\rm eq}^*) }/{\rho_{\beta}} \ll 1$. This is often quoted in various guises. 
Thinking of $\Da$ as the ratio of fluid density to solid, it is clear that for a gas-solid system $\Da \ll 1$ and the PSS is almost guaranteed to hold for a wide range of times and values of $\Tm$. For a liquid-solid system the densities may be similar and there is no guarantee $\Da \ll 1$. The applicability of the PSS model to liquid-solid systems has been previously discussed in \cite{Bisc63,gbor2004critical,Wen68}.

If we neglect only terms of order $\Da$ we must be aware that retained terms are of a greater magnitude. Specifically, the reduction requires $\Tm \gg \Da$ if terms of order $\Tm={R^* k}/{D}$ are retained. The kinetic-controlled result, Eq. \eqref{KinCon}, therefore only holds for $\Da \ll \Tm \ll 1$
and sufficiently large times, such that $\Tm t = \ord{1}$ or, in dimensional form, $t^* = \ord{\tau_g/\Tm}$ where $\tau_g/\Tm=\rho_{\beta}R^*/(k M_g  (c_{R,0}^*-c_{\rm eq}^*) )$. Since $s= \ord{1}$, the diffusion limited case holds for all time.

In reality the reduction to the two cases is somewhat  pointless - limitations are imposed through the requirement on $\Da$,  when \eqref{PSSSol} provides a simple expression without imposing extra restrictions.

%\label{subsec:Non-dimensionalisation}
%\begin{table}
%\begin{center}
%\begin{tabular}{ |c|c|c| } 
%\hline
%Parameter & Values & meaning \\
%\hline
%$R$ & $1.5\times10^{-5}$ m & Radius of the particle \\ 
%\hline
%$C_R$ & $P / (R_g T)$ & Average concentration in the surroundings\\ 
%\hline
%$D$ & $10^{-7}$ m$^2$/s & Effective Diffusion coefficient\\
%\hline
%$M_g$ & $2.02 \times 10^{-3}$ kg/mol & Molar mass of the fluid\\
%\hline
%$k$ & $10^{-4}$ m/s & Reaction constant \\
%\hline
%$\rho_{\beta}$ & 8000 $\textrm{kg/m}^3$ & Density of the $\beta$ - phase\\
%\hline
%\end{tabular}
%\caption{\tim{I suggest Table around here, so we can use it to check validity for our problem - are these values for the hydrogen case?} Values of $D$ obtained from \cite{oliva2018hydrogen}. In \cite{wang2022simulation}, they have a distribution of radii with mean value of $\sim 13.5\times 10^{-5}$ m, while in \cite{oliva2018hydrogen} is $5\times 10^{-6}$ m. For $C_R$, we need to determine the experimental conditions. For instance, in \cite{wang2022simulation} they use a range for $T$ from 300K to 340K and $P=1-1.5$ Mpa. Mg is well established . In most references, $k$ units are inverse of time. In order to obtain units of $m/s$ one can multiply the radius of the particle \cite{song2018numerical} (table 3).  \cristian{New values from \cite{homma2005gas}}}
%\end{center}
%\end{table}

%%%%%%%%%%%%%%%%%%%%%%%%%%%%%%%%%%%%%%%%%%%%%%%%%%%
\section{Approximate analytical solutions}
\label{sec:analytical_sols}

In this section, we derive approximate analytical solutions to the system~\eqref{nondim1}--\eqref{nondim2}. First, we obtain a correction to the pseudo-steady-state solution by constructing a perturbation expansion of the concentration in powers of $\Da$. We then analyze the behaviour of the system at early times and derive explicit expressions for both the concentration and the position of the reacting front in the small-time limit.

\subsection{Perturbation solution for small $\Da$}

We now apply a mathematically rigorous technique to define solutions to the system, with quantifiable errors. Specifically we will look for a solution through perturbation techniques, based on the assumption $\Da\ll 1$. 

We first seek a perturbation solution for the concentration in powers of $\Da$,
\begin{align}
   c(r,t)&=c_0(r,t) + \Da c_1(r,t) + \cdots
\end{align}

At leading order in $\Da$ the concentration satisfies
\begin{align}
\label{LO1}
0  = \frac{1}{r^2}\frac{\partial}{\partial r}\left(r^2\frac{\partial c_0}{\partial r}
\right)\,,    \,
\end{align}
subject to
\begin{align}
\label{LO2}
c_0(1,t) =c_R (t)\,, \qquad 
  \frac{\partial c_0(s,t)}{\partial r}   =  \Tm \, c_0(s,t)\,  \, .
\end{align}
This is the exactly the PSS model solved in the previous section. Consequently $c_0(r,t)$ is defined by equation \eqref{PSSC}.

The first order concentration is defined by
\begin{align}
\label{eq:FirstOrder}
\pad{c_0}{t}  = \frac{1}{r^2}\frac{\partial}{\partial r}\left(r^2\frac{\partial c_1}{\partial r}
\right)\,, \qquad 
c_1(1,t) = 0 \,,\qquad\frac{\partial c_1(s,t)}{\partial r} =  \Tm c_1(s,t)  \, .
\end{align}
This has solution
\begin{align}
\begin{split}
    c_1(r,t)&= \frac{1}{6}\frac{dB}{dt}\Bigg[(r-1)(r-2) +\\
    & \frac{s^2}{\Tm(\Tm^{-1}+ s - s^2)}\left(\Tm (s-1)(s-2)-(2s-3)\right)\left(1-\frac{1}{r}\right)\Bigg]
    \end{split}
\end{align}
where 
\begin{align}
    \nd{B}{t}= \frac{ s (2\Tm^{-1} +s)}{(\Tm^{-1} +s- s^2)^2}\, \frac{ds}{dt} \, .
\end{align}

The boundary position may now be calculated via
\begin{align}
    \nd s t = - \Tm \left(c_0(s,t) + \Da c_1(s,t) \right) = - \left. \frac{\partial c_0}{\partial r} \right|_{r = s} - \Da \left. \frac{\partial c_1}{\partial r} \right|_{r = s}   \,
\end{align}
where
\begin{align}
    \left. \frac{\partial c_0}{\partial r} \right|_{r = s} & = \frac{B}{s^2} \\
    \left. \frac{\partial c_1}{\partial r} \right|_{r = s} &= \frac{(1-s)^3}{3 (\Tm^{-1} +s-s^2)} \nd B t
\end{align}
From this we obtain 
\begin{align}
\frac{ds}{dt}=-\frac{1}{\Tm^{-1}+ s- s^2 + \frac{\Da }{3}\frac{s(1-s)^3(2\Tm^{-1}+ s)}{(\Tm^{-1}+ s- s^2)^2}}\,
\end{align}
and, after applying $s(0)=1$ we find

\begin{align}\label{pertsol_s}
t& = \Tm^{-1}(1-s)+\frac{1-s^2}{2} - \frac{1-s^3}{3}  \nonumber \\ & \hspace{1.0cm} + \frac{\Da }{6}
\left[
(1-s)\left(1-4\Tm^{-1}-s - \frac{2 \Tm^{-2}}{\Tm^{-1}+ s- s^2} \right) \right. \\
& + \left. \frac{12\Tm^{-2}}{\sqrt{1+4\Tm^{-1}}}
\left(\text{Arctanh}\left( \frac{1-2s}{\sqrt{1+4\Tm^{-1}}}\right)+\text{Arctanh}\left( \frac{1}{\sqrt{1+4\Tm^{-1}}}\right)
\right)
\right]
\,.\nonumber
\end{align}

It is worth noting that the appearance of $\Da$ in the perturbation solution is physically meaningful, as it introduces more of the underlying parameters governing the process, leading to a more accurate and interpretable description. In cases where some of these parameters are unknown, the analytical expression~\eqref{pertsol_s} can be employed in a fitting procedure to estimate them from experimental data, in a similar fashion to how the PSS approximation is widely applied. 

Now, the time until all of the $\alpha$-phase is used up, $t_f$, becomes
\begin{align}
    t_f =  \frac{1}{6} \left[ 1+\Da\left( 1-6\Tm ^{-1}+\frac{24 \Tm^{-2}}{\sqrt{1+4\Tm^{-1}}} \text{Arctanh} \left( \frac{1}{\sqrt{1+4\Tm^{-1}}} \right)\right)\right]+ \Tm^{-1} \,, 
\end{align}
which includes a correction term proportional to $\Da$ in the $t_f$ expression \eqref{final_t_PSS} from the PSS solution.  

Similar to the PSS case, an expression for the kinetic-controlled regime can be obtained by assuming $\Tm \ll 1$ in the perturbation solution~\eqref{pertsol_s}. However, this gives $s\approx 1-\Tm t$ as in the PSS case. In this case, the final time can be approximated as 
\begin{align}
   t_f = \Tm^{-1}\,.
\end{align}
In the diffusion-controlled limit, $\Tm \gg 1$, the interface position can be cast as
\begin{equation}\label{kinetic_1st_order}
t\approx\frac{1-s^2}{2}-\frac{1-s^3}{3}+\frac{\Da}{6}(1-s)^2\,,\end{equation}
while the final time is
\begin{align}
    t_f =\frac{1}{6}\left(1+\Da\right)\,.
\end{align}

%%%%%%%%%%%%%%%%%%%%%%%%%%%%%%%%%%%%%%%%%%%%%%%%%%%%

\subsection{Small time solution}
\label{subsec:sall_time_solution}

As discussed in Section~\ref{PSSmodel}, the PSS model is based on neglecting the time derivative in the diffusion equation, based on the assumption $\Da \ll 1$. If we consider a very early time-scale $\tau = \ord{\Da}$ or smaller then the PSS model no longer holds.  In this section, we develop an approximate analytical solution valid in the small-time regime, enabling us to better understand the system's initial transient behaviour and provide an initial condition for a numerical solution.

In order to study the small time behaviour of \eqref{nondim1}-\eqref{nondim4}, we first transform the diffusion equation \eqref{nondim1} into its Cartesian-like form using the change of variables $u(r, t) \equiv r c(r,t)$ and immobilise the moving boundary introducing the  space variable $\eta(r, t) \equiv (r-s)/(1-s)$. Now, the problem \eqref{nondim1}-\eqref{nondim4} reads 
\begin{align}
Da\, \left[	\frac{\partial u}{\partial \eta}\frac{\partial \eta}{\partial t}+\frac{\partial u}{\partial t}\right]&=\frac{1}{(1-s)^2}\frac{\partial^2 u}{\partial\eta^2}\,,
\label{planar_fixed1}\\
\left. u\right|_{\eta=1}&=1\,.\label{planar_fixed2}
\\
\frac{s}{1-s}\left.\frac{\partial u}{\partial \eta}\right|_{\eta = 0}&=\left(1+\Tm s \right)u|_{\eta = 0}\,,\label{planar_fixed3}
\\
\frac{ds}{dt}&=- \Tm \frac{u|_{ \eta =0}}{s}\,, \quad s(0)=1  \,. \label{planar_fixed4}
\end{align}
	
We are interested in the very early stages of the process. Thus, we redefine time through an arbitrary small parameter $\delta$ as $t \equiv \delta \tau$, which transforms \eqref{planar_fixed1}-\eqref{planar_fixed4} into 
\begin{equation}
\label{eq:ST-DiffEq}
\Da\left[	\frac{\partial u}{\partial \eta}\frac{\partial \eta}{\partial \tau}+\frac{\partial u}{\partial \tau}\right]=\frac{\delta}{(1-s)^2}\frac{\partial^2 u}{\partial\eta^2}\,,
\end{equation}
\begin{equation}
\label{eq:ST-BoundaryCond1}
 \left. u\right|_{\eta=1}=1\,.
\end{equation}
\begin{equation}
    \label{eq:ST-BoundaryCond2}
    \frac{s}{1-s}\left.\frac{\partial u}{\partial \eta}\right|_{\eta = 0}=\left(1+\Tm s \right)u|_{\eta = 0}\,,
\end{equation}
\begin{equation}
    \label{eq:ST-FreeBoundary}
    \frac{ds}{d\tau}=-\delta \Tm \frac{u|_{ \eta =0}}{s}  \,,
\end{equation}
As only a short period of time has elapsed, the interface is not far away from $R^*$, the radius of the particle. Therefore, it is useful to propose an ansatz as
\begin{equation}
\label{eq:ST-ansat}
s \approx 1-\lambda t^\alpha=1-\lambda \delta^\alpha\tau^\alpha\,,
\end{equation} 
where $\lambda$ and $\alpha$ constants that are yet to be adjusted.
Using the ansatz, we can determine the value of the scaling power $\alpha$:
\begin{equation}
\label{eq:ST-matchingscale}
-\lambda\alpha \delta^\alpha \tau^{\alpha -1}\approx-\delta \Tm u|_{\eta = 0}\,.
\end{equation}
Since $u|_{\eta = 0}\sim \mathcal{O}(1)$ for early times, we  conclude that $\alpha = 1$ and consequently
$\lambda = \Tm$ and 
\begin{equation} \label{s_small_t}
    s(t) = 1- \Tm t +\mathcal{O}(\delta^2)\,.
\end{equation}
This solution coincides with the kinetic-controlled result \eqref{KinCon}.

Equipped with these results, we can now study  the diffusion equation. Multiplying the whole equation by $(1-s)^2$ and taking into account that
	\begin{equation}
		\frac{\partial \eta}{\partial \tau} = -\frac{ds}{d\tau}\frac{1-r}{(1-s)^2}= \lambda \delta \frac{1-r}{(1-s)^2}\,,
	\end{equation}
	equation \eqref{eq:ST-DiffEq} becomes 
	\begin{equation}
    \label{unn}
		\Da \left(\lambda^2 \delta^2\frac{\partial u}{\partial \tau}+\lambda \delta (1-r)\frac{\partial u}{\partial \eta}\right) = \frac{\partial^2 u}{\partial \eta^2}\,.
	\end{equation}
	Allowing $\delta$ to become arbitrarily small equation \eqref{unn} reduces to 
	\begin{equation}
		 \frac{\partial^2 u}{\partial \eta^2} = 0\, ,
	\end{equation}
	and consequently $u$ is linear in $\eta$ at early times,
	\begin{equation}
		u(\eta, \tau) \approx A(\tau)+B(\tau)\eta\,,
	\end{equation}
	for, as yet, undetermined functions of time $A$, $B$. Applying the boundary conditions   \eqref{eq:ST-BoundaryCond1},  \eqref{eq:ST-BoundaryCond2}   leads to
		\begin{equation}\label{u_small_t}
			u(\eta,\tau) = 1-\frac{(1+\Tm s)(1-s)(1-\eta)}{s+(1+\Tm s)(1-s)}+\mathcal{O}(\delta^2)
		\end{equation}

Finally, for early times, the approximate expression for concentration in terms of the original variables $c(r,t)$ and $r$ is 
\begin{equation}
\label{eq:smalltimeconc}
	c(r,t)=\frac{1}{r}-\frac{1}{r}\frac{(1+\Tm s)(1-r)}{s }+\mathcal{O}(\delta^2)\,,
\end{equation}
%%
%\begin{equation}\label{eq:SmallTime}
%	s(t)=1 -t+\mathcal{O}(\delta^2)\,.
%\end{equation}
%%
{Note that in the denominator of \eqref{eq:smalltimeconc}, we have neglected the term proportional to $(1-s)$. Reintroducing it would only introduce a correction of order  $\delta^2$, which is of the same order as other neglected terms.}

An interesting point is that although the PSS does not hold at small times, the small time behaviour matches the kinetic-controlled solution (even in diffusion dominated situations). This means that, even though the kinetic solution does not hold for early times, according to the restrictions of the PSS approximation, it still captures the correct behaviour. 
This move away from the diffusion form may be inferred from the solution \eqref{PSSSol} by setting $s= 1-\varepsilon f(t)$, where $\varepsilon \ll 1$
\begin{align}
    t & =  \frac{s^3-1}{3} - \frac{s^2-1}{2} -  \Tm^{-1} (s -1)  =  - \frac{3 \varepsilon f}{3} + \frac{2 \varepsilon f }{2} -  \Tm^{-1} \varepsilon f + \mathcal{O}({\varepsilon^2})  \, 
\end{align}
and so, as $s \ra 1$ the solution $t \ra \Tm^{-1} (1-s)$ approaches the kinetic dominated result.

%%%%%%%%%%%%%%%%%%%%%%%%%%%%%
\section{Numerical solution}
\label{sec:numerics}

In this section, we present a finite difference scheme for solving the model. The numerical method is based on the transformed system \eqref{planar_fixed1}–\eqref{planar_fixed4}, which offers the advantage of solving the equations on a fixed unit domain $\eta \in [0,1]$, rather than on the evolving domain $r \in [s(t), 1]$ as in the original formulation \eqref{nondim1}–\eqref{nondim4}. This transformation not only simplifies the computational implementation but also enables us to use the results of the small-time analysis from Section~\ref{subsec:sall_time_solution} to initialize the scheme.

We apply the semi-implicit finite difference scheme for Stefan problems discussed in \cite{Caldwell,MITCHELL20091609}, which involves solving explicitly for $s(t)$ and implicitly for $u(\eta,t)$. Using second order central differences for the space derivatives and first order forward differences for the time derivatives, we obtain 
\begin{align}\label{FD_scheme}
&\left[\nu_1\Da^{-1} - \nu_2(1 - \eta_i)(1 - s^n)\left(\frac{ds}{dt}\right)^n\right] u_{i-1}^{n+1} \notag \\
&\quad - \left[2\nu_1\Da^{-1} + (1 - s^n)\right] u_i^{n+1} \notag \\
&\quad + \left[\nu_1\Da^{-1} + \nu_2(1 - \eta_i)(1 - s^n)\left(\frac{ds}{dt}\right)^n\right] u_{i+1}^{n+1} 
= -(1 - s^n) u_i^n\,,
\end{align}
where $\nu_1 = \Delta t/\Delta \eta^2$ and $\nu_2 = \Delta t/(2\Delta \eta)$. Equation \eqref{FD_scheme} holds for $i =1,2,\ldots,I-1 $ and $n = 0,1, 2, \ldots$, where $\Delta t$ and $\Delta \eta = 1/I$ denote the sizes of the temporal and spatial steps, respectively. The position of the moving front $s$ is updated using the Stefan condition \eqref{planar_fixed4}
\begin{equation}
s^{n+1} = s^n + \Delta t \left(\frac{ds}{dt}\right)^n\, \qquad \text{where} \qquad \left(\frac{ds}{dt}\right)^n = -\Tm\,\frac{u_1^n}{s}\,. 
\end{equation}
The boundary conditions \eqref{planar_fixed2}-\eqref{planar_fixed3} become 
\begin{align}
u_I^{n+1} &= 1\,,\\ 
u_1^{n+1} &= \frac{p}{1+3p}(-u_3^{n+1}+4u_2^{n+1})\quad \text{with}\quad p = \frac{s^n}{2\Delta \eta (1-s^n) (1+\Tm s^n) }\,,
\end{align}
where we have used a second order one-sided finite difference to discretise the space derivative in \eqref{planar_fixed3}. Finally, using the small time solution \eqref{u_small_t}-\eqref{s_small_t}, we arrive at 
\begin{align}
s^0 &= 1-\Tm\,\Delta t\,,\\ 
u^0_i &=  1-\frac{(1+\Tm s^0)(1-s^0)(1-\eta_i)}{s^0+(1+\Tm s^0)(1-s^0)}\,.
\end{align}

The semi-implicit scheme  \eqref{FD_scheme} can be formulated as a matrix linear system which can be solved by inverting the matrix of the system at each time-step. All numerical solutions were obtained in a uniform mesh of $10^{3}$ points, using a time-step size of $\Delta t = 4\times10^{-6}$. The implementation was carried out using a custom MATLAB code \cite{MATLAB}.

%%%%%%%%%%%%%%%%%%%%%%%%%%%%%%%%%%
\section{Results and discussion}
\label{subsec:DiffSolutions}

In this section, we first compare the approximate analytical solutions presented in Section~\ref{sec:analytical_sols} with the numerical results from Section~\ref{sec:numerics}, focusing on the solution behaviour across different limits of the dimensionless parameters $\Da$ and $\Tm$. We then demonstrate how the analytical solutions of the SCM found can be used to fit experimental data for the purpose of estimating unknown physical parameters of the process.

\subsection{Discussion of solutions and limiting behaviours}

In typical SCM-based experiments, measurements are used to determine the reacted fraction $X(t)$, which, as shown in~\eqref{reac_frac}, depends on the position of the reaction front $s(t)$. Thus, the evolution of $X(t)$ is directly governed by the behaviour of $s(t)$. 

\begin{figure*}
\centering
\includegraphics[width=\textwidth]{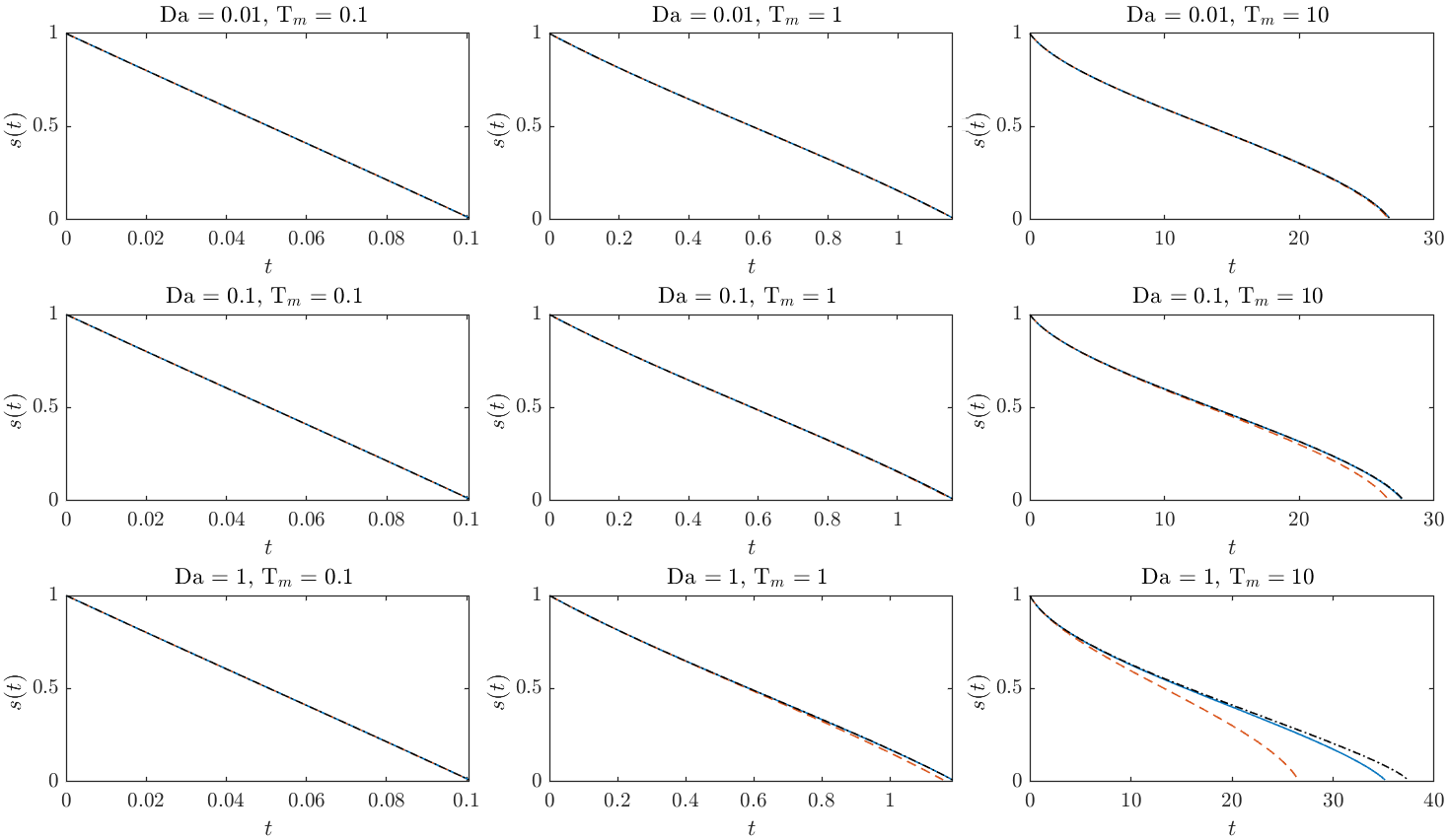}
\caption{Evolution of the moving front for different values of $\Da$ and $\Tm$. The solid line represents the numerical solution developed in Section~\ref{sec:numerics}, the dashed line corresponds to the pseudo-steady-state solution \eqref{PSSSol}, and the dash-dotted line shows the perturbation solution \eqref{pertsol_s}. }
\label{results1}
\end{figure*}

In Figure~\ref{results1} we show the evolution of $s(t)$ for various combinations of $\Da = 0.01$, $0.1$, $1$, and $\Tm = 0.1$, $1$, $10$, as predicted by the numerical solution, the PSS approximation~\eqref{PSSSol}, and the first-order perturbation solution~\eqref{pertsol_s}. The values of $\Da$ are kept below 1, as this range reflects the most physically realistic scenarios, corresponding to typical fluid-to-solid density ratios, as discussed in Section~\ref{subsec:nondim}. Then, the values $\Tm = 0.1$ to $\Tm = 10$ are chosen to study the diffusion and kinetic-controlled regimes, respectively, which are the ones typically discussed in the literature. 

Focusing on the comparison between the numerical and approximate analytical solutions, we observe that the discrepancies increase with larger values of $\Da$. This behaviour is expected: the PSS solution is derived in the limit $\Da = 0$, and the first-order perturbation solution is based on an expansion for $\Da \ll 1$. Therefore, the smaller the value of $\Da$, the closer the approximate solutions are to the full numerical solution, which retains all terms in the model. This trend is especially noticeable in the cases with $\Tm = 10$ (right panels of Figure~\ref{results1}). 

The superiority of the perturbation solution over the PSS approximation is also evident. The perturbation solution closely matches the numerical results in all cases, except for $\Da = 1$, $\Tm = 10$. Even in this most challenging case, where the PSS solution significantly deviates from the numerical one, the perturbation solution remains remarkably accurate up to $s \approx 0.25$, beyond which it begins to slightly diverge. This level of agreement is noteworthy given that $\Da = 1$ lies well outside the regime of validity for the perturbation approach, which formally requires $\Da \ll 1$. This confirms that the PSS approximation is valid primarily for gas--solid reactions ($\Da \ll 1$), while the perturbation solution provides more accurate results for gas-solid systems and remains reasonably accurate for liquid-solid reactions, where $\Da$ is closer to unity.

An interesting feature is the distinct change in the behaviour of $s(t)$ around $\Tm = 1$, which marks the transition between the diffusion-controlled ($\Tm \gg 1$) and kinetic-controlled ($\Tm \ll 1$) regimes. In particular, the kinetic-controlled regime exhibits a virtually linear evolution of $s(t)$ (left panels of Figure~\ref{results1}), consistent with the linear expressions obtained in~\eqref{KinCon} and~\eqref{kinetic_1st_order} for the PSS and perturbation solutions, respectively, in the limit $\Tm \ll 1$. The trends observed for large $\Tm$ are characteristic of the classical Stefan problem in spherical geometry (see, e.g.,~\cite{Font2013,Hill1987,McCue2008}). This is expected, as setting $\Tm^{-1} = 0$ in equations~\eqref{nondim1}–\eqref{nondim4}, previously rewriting~\eqref{nondim4} in terms of the gradient of $c$ using~\eqref{nondim3}, reduces the SCM to the classical one-phase Stefan problem for a sphere.

\begin{figure}\centering
\includegraphics[width=0.49\textwidth]{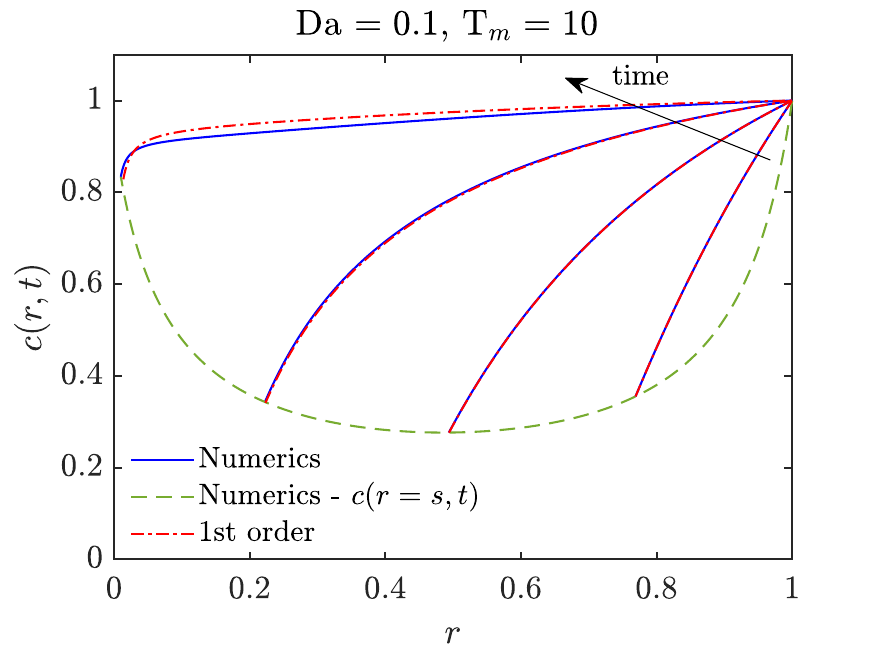}\includegraphics[width=0.49\textwidth]{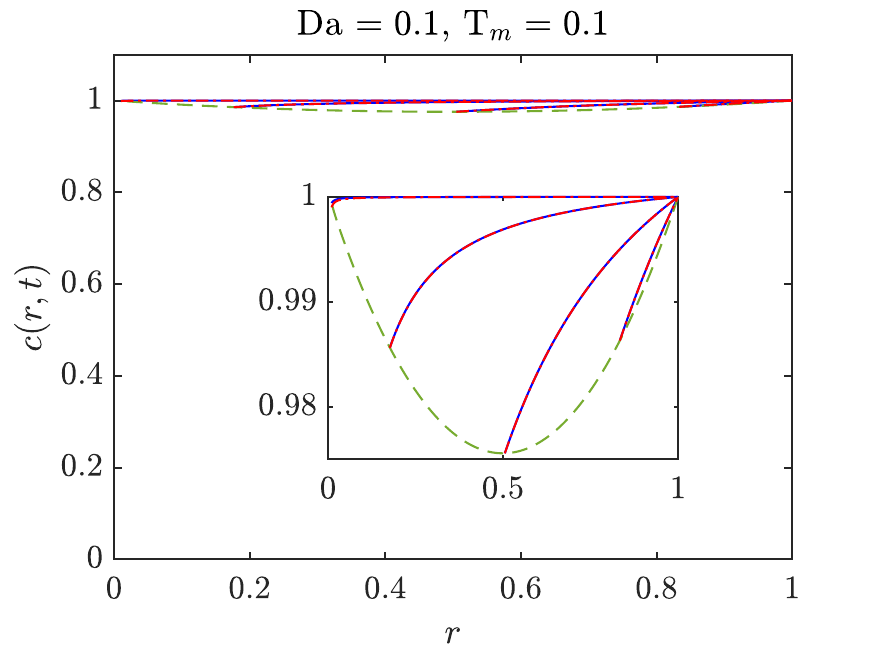}
\caption{Concentration profiles at four different times for the case $\Da=0.1$, $\Tm=10$ (left) and $\Da=0.1$, $\Tm=0.1$ (right). The inset shows a zoom in of the region near $c= 1$ for the case $\Da=0.1$, $\Tm=0.1$. The solid lines and dashed-dotted lines correspond to the numerical and first order perturbation solution, respectively. The dashed lines indicate the evolution of the concentration at the moving boundary, $c(r=s(t),t)$, from the numerical solution.}
\label{results2}
\end{figure}

In Figure~\ref{results2} we present the concentration profiles obtained for $\Da=0.1$, $\Tm=10$ (left panel) and $\Da=0.1$, $\Tm=0.1$ (right panel) at four different times during the reaction process (note these correspond to the center right and center left solutions for $s(t)$ in Figure~\ref{results1}). We show the first order perturbation solution (dash-dotted lines) along with the numerical solution (solid line), and omit those of the PSS since they are less accurate, as previously discussed. The dashed-line corresponds to the evolution of the concentration at the reacting boundary. The case with $\Da = 0.1$ and $\Tm = 10$, corresponding to the diffusion-limited regime, exhibits significant concentration variation within the reacted layer. In this scenario, the reaction proceeds much faster than diffusion, leading to depletion of the concentration near the reaction front $s(t)$. This occurs because diffusion is too slow to replenish the consumed gas at the reacting boundary, $s(t)$. 

The case with $\Da = 0.1$, $\Tm = 0.1$, corresponding to the kinetic-limited regime, is characterized by concentrations remaining close to 1 throughout the domain. This reflects the fact that diffusion is much faster than reaction, allowing the concentration near the moving boundary to rapidly equilibrate with the external concentration due to efficient mass transport. The inset plot for $\Da = 0.1$, $\Tm = 0.1$ provides a close-up view of the concentration profiles near $c \approx 1$, revealing shapes qualitatively similar to those in the diffusion-limited regime, though with concentration variations that are an order of magnitude smaller. Finally, we note the excellent agreement between the numerical solution and the first-order perturbation approximation, with only minor discrepancies appearing near the end of the process in the diffusion-limited case.

\subsection{Comparison to synthetic data}

Experimental studies often employ fitting procedures to determine unknown physical parameters of the process. In solid-gas reaction experiments using the SCM, typical unknowns include the reaction rate constant $k$, the diffusion coefficient of the ash layer $D$ or the equilibrium concentration $c_{\rm eq}^*$. The experimental data commonly consist of the reacted fraction $X$ measured as a function of time, $t^*$. Therefore, the fitting process generally involves estimating the unknown parameters by fitting datasets of the form $(t_{\text{exp},i}, X_{\text{exp},i})_{i \in \{1,\dots,N\}}$. In this section, we show how the small time limit \eqref{s_small_t} and the perturbation solution \eqref{pertsol_s} can be used to determine the parameters $k$, $D$ and $c_{\rm eq}^*$ from an experimental dataset. In addition, the PSS approximation \eqref{PSSSol} will also be tested as it is widely used in the literature. 

\begin{figure}\centering
\includegraphics[width=0.49\textwidth]{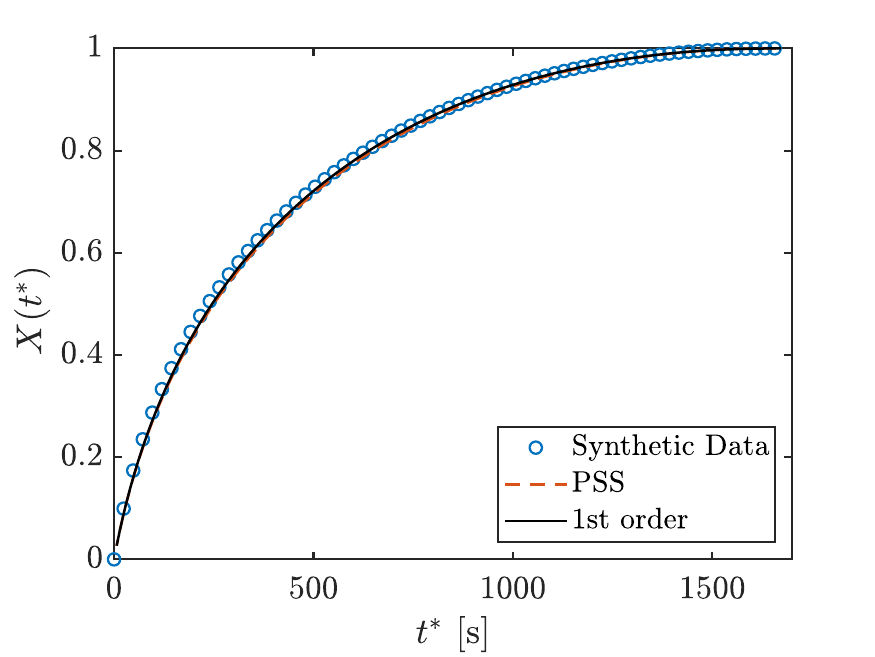}
\includegraphics[width=0.49\textwidth]{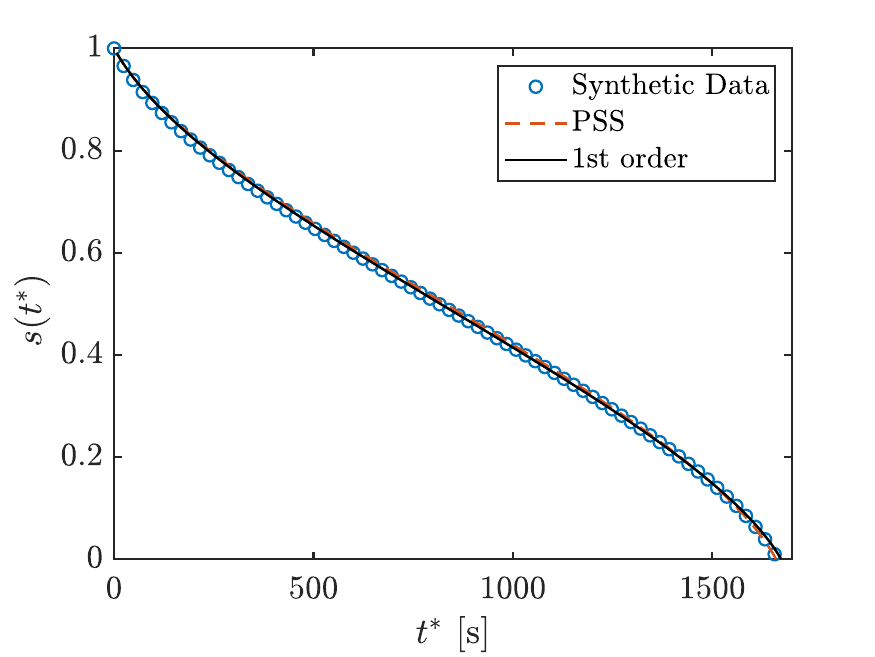}
\includegraphics[width=0.49\textwidth]{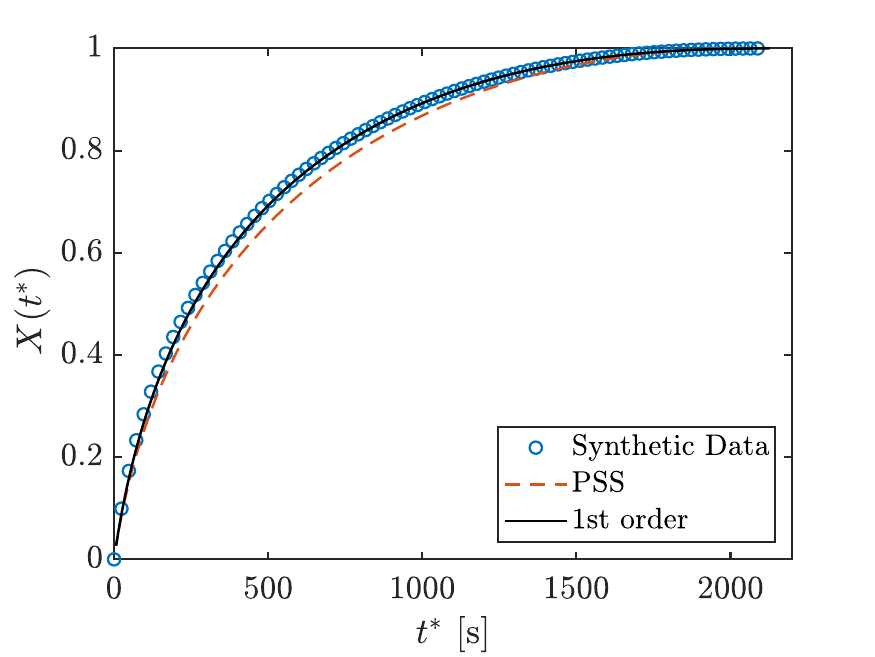}
\includegraphics[width=0.49\textwidth]{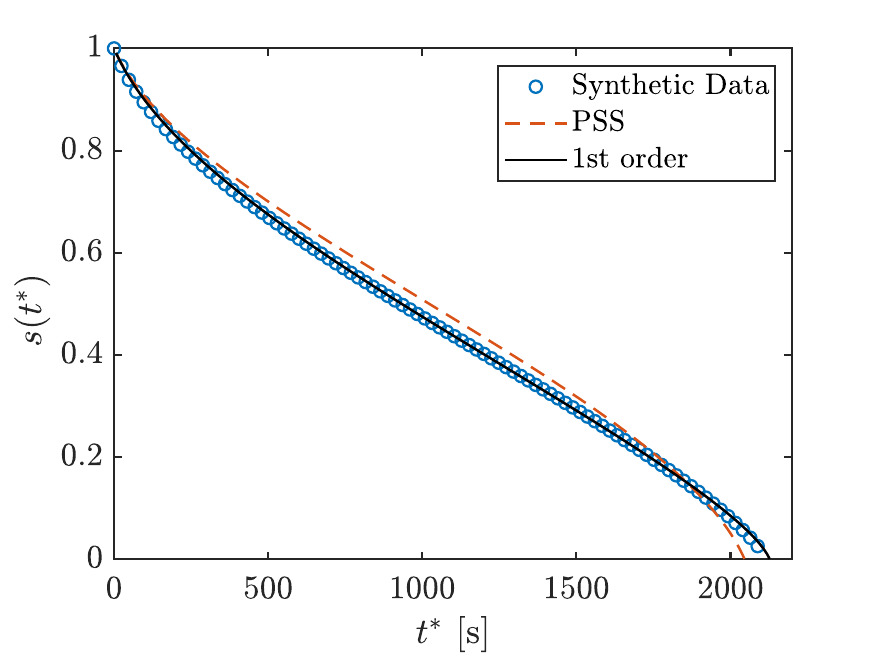}
\caption{Prediction of the reacted fraction $X(t^*)$ and position of the moving front $s(t^*)$ using the parameter values obtained in the fitting process for Dataset 1 (Top panels) and Dataset 2 (Bottom panels). }
\label{hey}
\end{figure}

We generate two synthetic datasets by numerically solving the full model \eqref{nondim1}-\eqref{nondim4} with the numerical scheme developed in \ref{sec:numerics}. The first data set corresponds to the numerical solution of the model for $\Da=0.1$, $\Tm = 10$ and the second data set for $\Da=1$, $\Tm = 10$. In order to present the dataset in dimensional time, we set the characteristic time scale to $\tau = \tau_{g} = 100$ min. To avoid a dataset with an unrealistic number of data points, we only take 70 time points from the numerical solution, i.e. $N=70$. Since our solutions are in terms of the interface position, we will proceed with the fitting of \(s_{\text{exp},i} = (1 - X_{\text{exp},i})^{1/3}\) rather than $X_{\text{exp},i}$. The evolution of $s_{\text{exp},i}$ and $X_{\text{exp},i}$ for the two datasets is shown with open circles in Figure~\ref{hey}.

The fitting process is done in two steps. The first, involves finding a reference value of the parameter group $\tau_{\text{eff}}\equiv\tau/\Tm$ using the dimensional form of the small time solution \eqref{s_small_t}, i.e. $s(t^*) = 1-(\Tm/\tau)\,t^* = 1-(t^*/\tau_{\text{eff}}) $. Since the small time limits are short-lived, we take the first two points of the dataset to compute the slope $1/\tau_{\text{eff}}$ of the straight line connecting them. The second, involves using \eqref{pertsol_s} to find the values of $\Tm$ and $\Da$, that best minimize the quadratic error   
\begin{align}\label{error}
    E_{Q}= \sum_{i=3}^{N}\left(s_{\text{exp},i}-s_{ \theta,i}\right)^2\,,
\end{align}
where $\theta$ corresponds to $\theta=\left(\Tm,\Da\right)$ in the case of the first-order perturbation solution, and $\theta=\Tm$ for the PSS. In the minimisation of \eqref{error}, expression \eqref{pertsol_s} is used in dimensional time. Hence, the left hand side of \eqref{pertsol_s} becomes $t_s^*/\tau = t_s^*\,\Tm/\tau_{\text{eff}}$. In this way, expression \eqref{pertsol_s} can be rearranged such that the value $\tau_{\text{eff}}$, found in the first step, can be used in the second step, and the only remaining fitting parameters become $\Tm$ and $\Da$. The minimization is implemented using MATLAB \textit{fminsearch} function. 

\begin{table}
\begin{center}
\begin{tabular}{|c|c|c|} 
\hline
\textbf{1st Order} & \textbf{Dataset 1}  & \textbf{Dataset 2} \\
\hline
Error & 0.5$\times10^{-3}$  & 0.6$\times10^{-3}$ \\
\hline 
Fitted $\Tm$ & 7.77  & 8.17 \\ 
\hline
Fitted $\Da$ & 0.12  & 0.81 \\ 
\hline
\textbf{PSS} & \textbf{Dataset 1} & \textbf{Dataset 2} \\
\hline
Error &  2$\times10^{-3}$ & 5.12$\times10^{-2}$\\
\hline
Fitted $\Tm$ & 8.22  & 11.49  \\ 
\hline
Fitted $\Da$ & -     &  - \\ 
\hline
\end{tabular}
\caption{
Summary of the fitting process results for synthetic data generated with $\Da=0.1$, $\Tm = 10$, $\tau = 100$ min (Dataset 1) and $\Da=1$, $\Tm = 10$, $\tau = 100$ min (Dataset 2).} 
%The estimate for the time scale was $\tau = 700.6$ s for Dataset 1 and $\tau = 701.58$ s for Dataset 2.
\label{table_fitting}
\end{center}
\end{table}

The reference values for $\tau_{\text{eff}}$ obtained using the small time limit for our two synthetic datasets are $\tau_{\text{eff}} = 700.6$ s and $\tau_{\text{eff}} = 701.58$ s for Dataset 1 and 2, respectively. In Table~\ref{table_fitting} the results of the second step of the fitting process are presented. We observe that in Dataset 1, the error $E_Q$ obtained using the first-order perturbation solution is an order of magnitude smaller than that obtained with the PSS approximation. In Dataset 2, the error is reduced by two orders of magnitude when using the first-order perturbation solution compared to the PSS approximation. This highlights the advantage of employing the first-order perturbation solution~\eqref{pertsol_s} over the PSS for fitting experimental data. 

A second advantage of the first-order perturbation approach is its ability to provide an estimate of the Damköhler number, $\Da$, which the PSS solution cannot offer. The estimated values of $\Da$—0.12 for Dataset~1 and 0.81 for Dataset~2—are close to the true values used to generate the synthetic data (0.1 and 1.0, respectively). With regard to the Thiele modulus $\Tm$, both the first-order perturbation and PSS solutions yield reasonable approximations of the true value $\Tm = 10$ used in both datasets. The PSS result is slightly closer to the true value in the case of Dataset~1. 

In Figure~\ref{hey}, we present the predictions of $s(t^*)$ and $X(t^*)$ obtained from the fitting procedure using both the first-order perturbation and PSS solutions, plotted alongside the synthetic data. For the dataset corresponding to $\mathrm{Da} = 0.1$, both the first-order perturbation and PSS solutions closely match the data, and the improvement offered by the perturbation solution is not visually evident. In contrast, for the second dataset ($\Da = 1$), the first-order perturbation solution provides a visibly better fit than the PSS. This result is expected, since the first-order correction is of order $\mathcal{O}(\Da)$; as $\Da$ increases, the correction captures the behaviour of $s(t^*)$ and $X(t^*)$ more accurately than the PSS solution, which assumes $\Da = 0$.

The last step requires translating the results of the fitting procedure to values of physical process parameters. As mentioned previously, in solid-gas reaction systems, the most common unknowns are $k$, $D$ and $c_{\rm eq}^*$. Recalling the definitions of the fitting parameters in terms of the original physical quantities
$$\Da = \frac{M_g (c_R^* - c_{\rm eq}^*)}{\rho_{\beta}}, \quad \Tm = \frac{R^* k}{D}, \quad \tau = \Tm \,\tau_{\text{eff}} = \frac{\rho_{\beta} R^{*2}}{M_g (c_R^* - c_{\rm eq}^*) D}\,,$$
the unknown physical parameters of interest can be inferred from 
%$$c_{\rm eq}^* = c_R^* + \frac{\Da\rho_{\beta}}{M_g}\,, \quad D= \frac{\rho_{\beta} R^{*2}}{M_g (c_R^* - c_{\rm eq}^*) \Tm \,\tau_{\text{eff}}}\,, \quad k = \frac{\Tm D}{R^*}\,,$$ 
%or, alternatively,
$$c_{\rm eq}^* = c_R^* + \frac{\Da\rho_{\beta}}{M_g}\,, \quad 
D= \frac{R^{*2}}{\Da \Tm \,\tau_{\text{eff}}}\,, \quad 
k = \frac{R^*}{\Da \tau_{\text{eff}}}\,,$$ 
if the 1st order perturbation solution is used to fit the data. Although, as demonstrated, the PSS model tends to produce larger errors compared to the first-order correction during fitting, it may still yield approximate values for $k$ and $D$, provided that $c_{\rm eq}^*$ is either known or negligible relative to $c_R^*$. In such cases, $c_{\rm eq}^*$ can be omitted from the expressions for $\tau$ and $\Da$.

%%%%%%%%%%%%%%%%%%%%%%%%%%%
\section{Conclusions}
\label{sec:Conclusions}

In this work, we derived and analyzed the shrinking core model (SCM), which provides a mathematical framework for describing the evolution of a spherical solid particle undergoing reaction with a surrounding fluid. A careful non-dimensionalisation of the governing equations reveals that the widely used pseudo-steady-state approximation must be applied with caution, since it is generally valid only for solid-gas reactions, where the gas density is significantly lower than that of the solid. For solid-liquid systems, where the fluid and solid densities are comparable, the non-dimensional analysis indicates that transient effects may become significant. As a result, applying the pseudo-steady-state approximation in such cases may yield inaccurate predictions and unreliable parameter estimates when fitting experimental data. In addition, the non-dimensionalisation introduces a key parameter: the ratio of diffusion to reaction time scales. We show that the commonly cited \textit{kinetic limited} and \textit{diffusion limited} regimes naturally emerge as asymptotic limits of this parameter tending to zero and infinity, respectively, within the pseudo-steady-state framework.

We have derived approximate analytical solutions to the SCM using a perturbation method that introduces a first-order correction to the pseudo-steady-state solution. These perturbation and pseudo-steady-state solutions were then compared against the numerical solution of the full system. Our results demonstrate that the perturbation solution offers a clear improvement over the pseudo-steady-state approximation, particularly in the diffusion-limited regime. The numerical simulations also confirm that the pseudo-steady-state approximation breaks down when the fluid and solid densities become comparable, whereas the perturbation solution continues to provide reasonable accuracy under such conditions. 

The system's behaviour at early times is also analyzed analytically, and an explicit expression for the small-time solution is derived. This enables the design of a simple fitting procedure that combines the small-time solution with either the pseudo-steady-state or perturbation approximation. Using synthetic data, we demonstrate how the method can be applied to estimate physical parameters from experimental observations with minimal computational cost, offering a practical alternative to full numerical simulations for inferring unknown process parameters.

%%%%%%%%%%%%%%%%%%%%%%%%%%%
\section*{Acknowledgments}
This publication is part of the research project TED2021-131455A-I00 of the Spanish \emph{Agencia Estatal de Investigación} (AEI) of \emph{Ministerio de Ciencia, Innovaci\'on y Universidades} (MCIU), funding all authors. F. Font, T. Myers and R. Olwande, are also funded by PID2023-146332OB-C21 from AEI of MCIU. C. Moreno-Pulido is supported by PID2023-146332OB-C22 from AEI of MCIU. F. Font gratefully acknowledges the SGR programme (Ref. 2021-SGR-01045) and the \emph{Serra-Hunter} Programme of the \emph{Generalitat de Catalunya}.

%%%%%%%%%%%%%%%%%%%%%%%%%%%
\bibliographystyle{elsarticle-num-names} 
\bibliography{mybib}

\end{document}